\documentclass[aps,a4paper,amsmath,twocolumn,amssymb,titlepage]{revtex4-1}
\usepackage{here,wrapfig, amsmath, amssymb, braket, mathrsfs}
\usepackage{csquotes, amsfonts, multirow, tabularx, array, bm, txfonts, bibentry, ulem, color}
\usepackage{xcolor, etoolbox}
\usepackage{graphicx}

\usepackage{xparse, soul, xcolor}
\usepackage{lipsum}

\NewDocumentCommand{\sotwo}{O{red}O{black}+m}
    {%
        \begingroup
        \setulcolor{#1}%
        \setul{-.5ex}{.4pt}%
        \def\SOUL@uleverysyllable{%
            \rlap{%
                \color{#2}\the\SOUL@syllable
                \SOUL@setkern\SOUL@charkern}%
            \SOUL@ulunderline{%
                \phantom{\the\SOUL@syllable}}%
        }%
        \ul{#3}%
        \endgroup
    }

\newcommand{\sset}{{\bm \sigma}}
\newcommand{\hset}{{\bm h}}
\newcommand{\dset}{{\bm d}}
\newcommand{\up}{\triangle}
\newcommand{\down}{\bigtriangledown}

\begin{document} 

\title{Transforming Generalized Ising Model into Boltzmann Machine}
\author{Nobuyuki Yoshioka} 
\email{\tt nysocloud@g.ecc.u-tokyo.ac.jp}
\author{Yutaka Akagi, and Hosho Katsura}
\affiliation{Department of Physics, University of Tokyo, 7-3-1 Hongo, Bunkyo-ku, Tokyo 113-0033, Japan}

\begin{abstract}
We find an exact mapping from the generalized Ising models with many-spin interactions to equivalent Boltzmann machines, i.e., the models with only two-spin interactions between physical and auxiliary binary variables accompanied by local external fields.
More precisely, the appropriate combination of the algebraic transformations, namely the star-triangle and decoration-iteration transformations, allows one to express the model in terms of fewer-spin interactions at the expense of the degrees of freedom. 
Furthermore, the benefit of the mapping in Monte Carlo simulations is discussed. 
In particular, we demonstrate that the application of the method in conjunction with the Swendsen-Wang algorithm drastically reduces the critical slowing down in a model with two- and three-spin  interactions on the Kagom\'e lattice.
\end{abstract}
\maketitle

\section{Introduction}
Inspired by the tremendous success in machine learning fields such as the image or speech recognition, climate prediction, and market analysis,
the increasing number of studies have shown the validity of artificial neural network in physical problems.
For instance, physicists have recognized that the phase classification can be done in a parallel fashion to the ordinary machine learning tasks by using the data of the classical, quantum, or auxiliary degrees of freedom~\cite{Carrasquilla_Machine_2017, Wang_Discovering_2016, Torlai_Learning_2016, Ohtsuki_Deep_2016, Ohtsuki_Deep_2017, Yoshioka_Learning_2018, Nieuwenburg_Learning_2017, Chng_Machine_2017, Wetzel_Unsupervised_2017, Huembeli_Identifying_2018, Araki_Phase_2018, Rem_Identifying_2018}.
Another important direction focuses on the representability itself.
The applications such as the variational wave functions describing ground states turn out to outperform the state-of-the-art methods~\cite{Carleo_Solving_2017, Nomura_Restricted_2017, Glasser_Neural_2018, Carleo_Constructing_2018, Freitas_Neural_2018}, 
and are also valid in estimating the Boltzmann factor of a system to accelerate the classical Monte Carlo (MC) simulations~\cite{Torlai_Learning_2016, Huang_Accelerated_2017}.
Furthermore, numerous studies reveal the ability to precisely express nonlinear functions such as the wave functions of stabilizer states and chiral topological states \cite{Deng_Machine_2017, Deng_Quantum_2017, Gao_Efficient_2017, Chen_Equivalence_2018, Jia_Efficient_2018, Wang_Exploring_2017}.
This is not only insightful from the perspective of exploring soluble models but also fruitful in constructing new numerical techniques including optimization of variational functions, MC simulations, and so on.

Since the invention of the MC method, physicists have long made efforts to develop versatile and efficient simulation methods to investigate statistical models.
In classical lattice systems, the single-spin flip (SSF) algorithm is undoubtedly one of the most widely-used techniques, as it is model independent. 
The locality of the variables involved in a single update procedure, however, inevitably leads to a severe slowing down near critical points or at low temperature for non-ordering systems.
One of the solutions is to apply the global updates such as the cluster algorithms~\cite{Swendsen_Nonuniversal_1987, Wolff_Collective_1989}, worm algorithm~\cite{Prokofev_Worm_2001}, and loop algorithm~\cite{Evertz_Loop_2003}, but they are mostly restricted to two-body interacting systems.
Developing a generic technique applicable to a wide variety of systems involving many-body interactions is highly challenging.

In this work, we establish a mapping from the generalized Ising model to the Boltzmann machine (BM).
The former, which includes many-spin interactions, is used to describe magnetic and thermodynamic properties of solids, the effective model of alloys, spin glass models, and so on \cite{Biroli_Lattice_2001, Blum_Using_2005, Berthier_Theoretical_2011}.
The BM, on the other hand, is an expression of probability distribution by two-spin interactions between physical (visible) and auxiliary (hidden) binary degrees of freedom.
%\txtr{
The mapping procedure goes as follows: 
one first decomposes a $p$-spin Ising interaction
 into a sum of three- and $(p-1)$-spin interactions by adding an auxiliary spin.
 This is repeated for the latter many-spin interaction until the total interaction is expressed in terms of two- and three-spin interactions.
Finally, each three-spin interaction is transformed into two-spin interactions and single-spin terms.
The idea of such consecutive decomposition can also be seen in the context of quantum annealing~\cite{Biamonte08, Lechner15, Leib_Transmon_2016, Choi_2011}.
%}

After constructing %a
%\txtr{
the
%}
rigorous mapping 
%\txtrout{between these representations}
, we take advantage of it by presenting a novel global update scheme;
the application of the Swendsen-Wang (SW) algorithm to the exactly surrogate BM. 
By comparing the autocorrelation times of the visible spin configurations, we demonstrate in a model with two- and three-spin interactions on the Kagom\'e lattice that the sampling efficiency of the cluster algorithm on the BM at the critical temperature is drastically improved compared to that of SSF performed on the original Hamiltonian.
While the mapping introduced in this paper is applicable to the generalized Ising model in any number of dimensions, our results are presented in the above-mentioned model for  clarity and simplicity.

%----------------------------------------------------%
\begin{figure}[t]
\begin{center}
\begin{tabular}{c}
  \begin{minipage}{0.97\hsize}
    \begin{center}
     \resizebox{0.95\hsize}{!}{\includegraphics{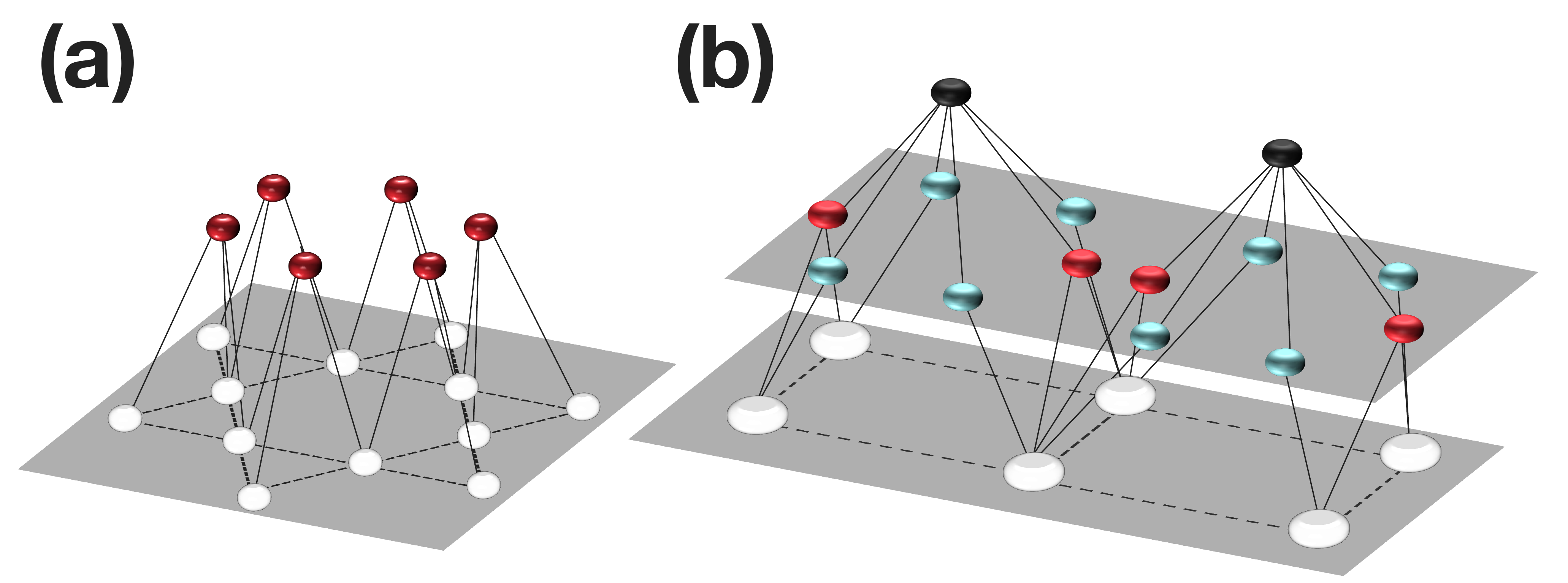}}

   \end{center}
  \end{minipage}
\end{tabular}
\end{center}
\caption{\label{RBMDBMGraphic} (Color Online) Schematic picture of (a) restricted Boltzmann machine (RBM) that is equivalent to a model with three-spin Ising-type interaction,  (b) deep Boltzmann machine (DBM) that is equivalent to a model with four-spin Ising type interaction. The white and black objects denote visible and deep spins, respectively. 
Also, the blue and red objects are the hidden spins introduced by the decoration-iteration transformation (DIT) and star-triangle transformation (STT), which are given by Eq. (\ref{DIT}) and Eq. (\ref{STT}). The presence of the layers is denoted by the gray planes. Note that the mapping to the RBM or DBM is applicable irrespective of the spatial dimension.}
\end{figure}
%----------------------------------------------------%

The remainder of this paper is organized as follows. 
In Sec. \ref{transformSection} we first introduce the most primary transformation techniques, namely the decoration-iteration and star-triangle transformations. Embedding the hidden spins by combining the two, an arbitrary many-spin interaction is shown to be mappable to the BM.
Furthermore, multiple interaction terms can be taken into account by simply considering the embedding procedures independently.
In Sec. \ref{MCSection}, we see that the transformation is numerically beneficial since the existing cluster update can be applied to the BM.
Finally, the summary for the current work and discussion concerning the future direction are given in Sec. \ref{DiscussionSection}.
For the completeness of the paper, in Appendix \ref{pspin2DBMSection} we derive the explicit expression of the BM mapped from the generalized Ising model. 
In Appendix \ref{GibbsSection}, we qualitatively see that the block Gibbs sampling for the RBM equivalent to the ferromagnetic Ising model is highly inefficient.
The partition function of the pure three-spin interacting model on the Kagom\'e lattice, from which the absence of the phase transition follows, is calculated in Appendix \ref{PartitionSection}.
Also, two schemes to perform the cluster update under magnetic field are given in Appendix \ref{MagClusterSection}, and the observation of the physical quantity in the extended space is discussed in Appendix \ref{ObservationSection}.

\section{Algebraic Transformation of Boltzmann Factors}\label{transformSection}
In this section, we find the equivalence of the generalized Ising models and the BMs with specific architectures, i.e., the restricted Boltzmann machine (RBM) and the deep Boltzmann machine (DBM). [See Fig. \ref{RBMDBMGraphic} for the graphic representation.]
First, we state the definition of the RBM and DBM, which give some probability distributions of visible spins with 
auxiliary binary degrees of freedom, or the hidden spins. 
The Boltzmann factor of an RBM is given as \cite{Smolensky_Information_1986, Hinton_Training_2002}
\begin{eqnarray}
\pi(\sset) &=& \sum_{\hset} \widetilde{\pi}(\sset, \hset),\\
\widetilde{\pi}(\sset, \hset ) &=& \exp\left\{\sum_{i,j}W_{ij}\sigma_i h_j  + \sum_i a_i\sigma_i + \sum_j b_j h_j \right\}, \label{RBMEqn}
\end{eqnarray}
where $\pi(\sset)$ and $\widetilde{\pi}(\sset, \hset)$ are the Boltzmann factor of $N_v$ visible spins and the RBM with additionally $N_h$ hidden spins, respectively. 
Here, $\sigma_i$ and $h_j$ are the $i$-th visible and $j$-th hidden spins that take either $+1$ or $-1$, coupled via the interaction $W_{ij}$.
The local external field is given as $a_i$ and $b_j$ for visible and hidden spins, respectively.
We denote the visible and hidden spin configurations by $\sset := (\sigma_1, \sigma_2, ..., \sigma_{N_v}) \in {\cal S} $ and $\hset:= (h_1, h_2, ..., h_{N_h}) \in {\cal H}$, where ${\cal S} = \{-1, 1\}^{N_v}$ and ${\cal H} = \{-1, 1\}^{N_h}$ are the sets of all possible binary spin configurations for visible and hidden spins, respectively. 
Also, to discriminate between the spaces with and without the hidden spins, we call $\cal S$ as the ``original space'' and ${\cal S} \cup {\cal H}$ as the ``extended space.''
Note that the absence of the intralayer couplings is reflected in the bipartite structure as is depicted in Fig. \ref{RBMDBMGraphic}(a).
The BM without any restriction on the connectivity is not considered in the following.

A crucial notion in introducing the DBM is the ``layer,'' for which we provide a description based on the graphical understanding of Fig. \ref{RBMDBMGraphic}.
As aforementioned and also obvious from Fig. \ref{RBMDBMGraphic}(a),  no coupling is present between the visible or hidden spins in the RBM. 
Such a bipartite structure is understood as a BM with the ``visible layer and one hidden layer.'' 
Namely, the visible layer contains all the visible spins, and the hidden layer consists of hidden spins that are interacting with one or more visible spins. 
We may analogously construct another hidden layer on top as depicted in Fig. \ref{RBMDBMGraphic}(b), which is a BM with the ``visible layer and two hidden layers'' in turn.
In principle, one may add as many hidden layers as desired.
Such an architecture is generally referred to as the DBM.
The maximum number of layers obtained by mapping from the generalized Ising model is two as we discuss in Sec. \ref{pspinSection},
and therefore we discriminate the first and second hidden layers as the ``hidden layer'' and ``deep layer'' in the following.
The spins included in these layers are correspondingly referred to as the ``hidden spins'' and ``deep spins.''
To be concrete,  the Boltzmann factor of a DBM is given as
\begin{eqnarray}
\pi(\sset) &=& \sum_{\hset,\dset} \widetilde{\pi}(\sset, \hset, \dset),\\
\widetilde{\pi}(\sset, \hset,\dset) &=& \exp\left\{\sum_{i,j}W_{ij}\sigma_i h_j  + \sum_{j,k} W'_{jk}h_j d_k + \right.\nonumber \\
&& \left. + \sum_i a_i\sigma_i + \sum_j b_j h_j  + \sum_k b'_k d_k\right\}, \label{DBMEqn}
\end{eqnarray}
where $\widetilde{\pi}(\sset, \hset, \dset)$ is the Boltzmann factor for each spin configuration, $d_k$ is the $k$-th deep spin with the local field $b'_k$, and $W'_{jk}$ is the interaction between the $j$-th hidden spin and the $k$-th deep spin. 
As is the case for the visible and hidden spins, a configuration of the deep spins is denoted by $\dset := (d_1, d_2, ..., d_{N_d}) \in {\cal D}$, where ${\cal D} = \{-1, 1\}^{N_d}$ is the set of all possible deep spin configurations.
The union ${\cal S} \cup {\cal H} \cup {\cal D}$ is also referred to as the extended space in the following.

The probabilistic network given by quadratic terms as Eqs. (\ref{RBMEqn}) and (\ref{DBMEqn}) is known to be powerful to express non-linear functions and in fact applied widely in the field of machine learning, condensed matter physics, quantum physics, and so on~\cite{Deng_Machine_2017, Deng_Quantum_2017, Gao_Efficient_2017, Chen_Equivalence_2018, Jia_Efficient_2018, Carleo_Solving_2017, Nomura_Restricted_2017, Glasser_Neural_2018}.
In the remainder of this section, we find the exact mapping from the generalized Ising model to the RBM or DBM.

\subsection{Transformation techniques}
\begin{figure}[t]
\begin{center}
\includegraphics[width=0.9\hsize]{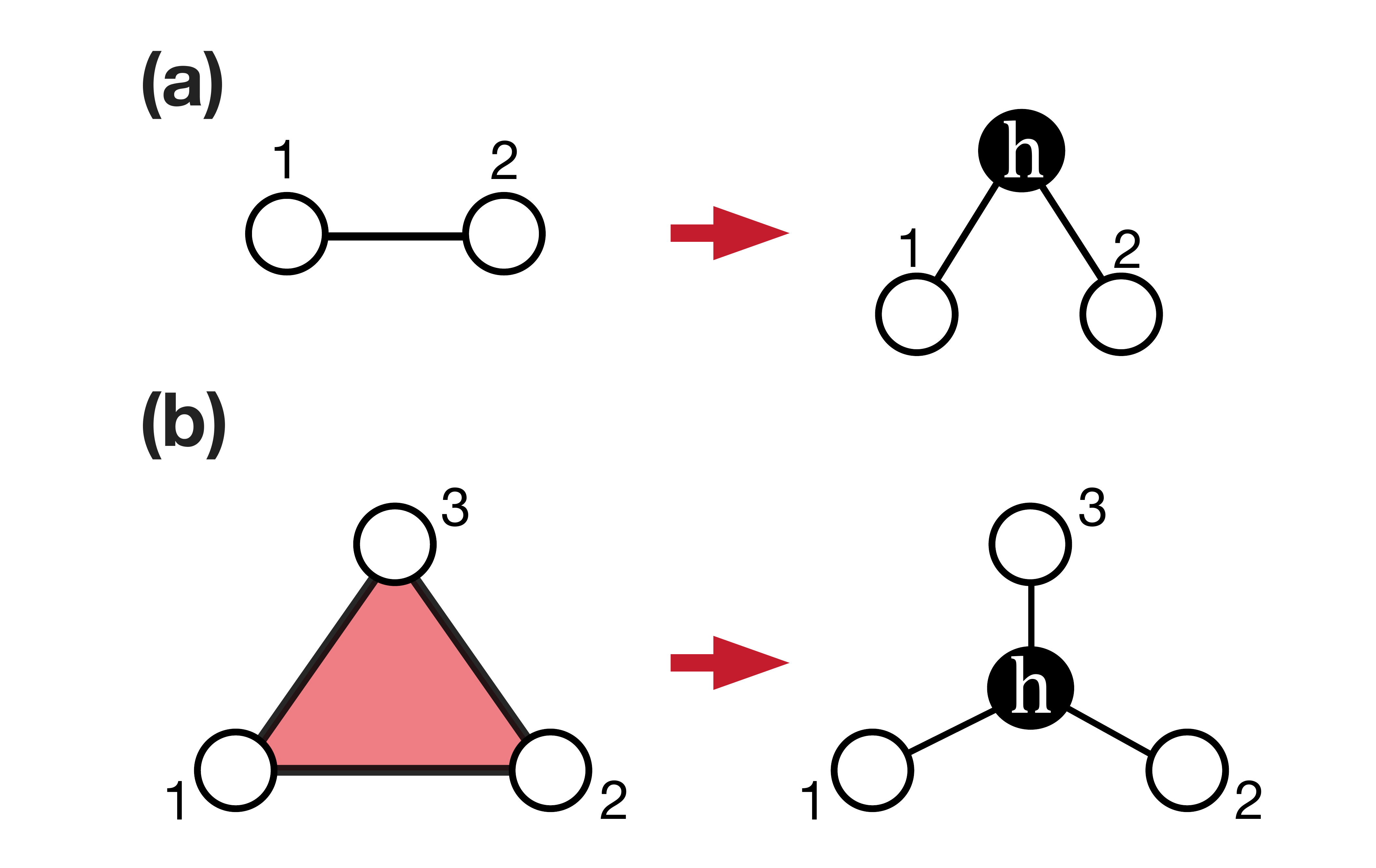}
\caption{(Color Online) Schematic description of the mapping techniques: (a) Decoration-iteration transformation and (b) star-triangle transformation. The white and black circles correspond the visible and hidden spins, respectively, and the numbers denote the labels of the visible spins. The solid lines and the filled region denote the two- and three-spin interactions, respectively.
%\txtb{[The font of h in the figure has been changed due to illustrator...]}
}
\label{transformation}
\end{center}
\end{figure}

We introduce two mapping techniques to embed hidden spins as is graphically described in Fig. \ref{transformation}: the decoration iteration transformation (DIT) and star-triangle transformation (STT). 
Note that the newly embedded spins are auxiliary, and the original interactions are realized by tracing out such degrees of freedom.

%----------------------------------------------------------------------------------------%
%----------------------------------------------------------------------------------------%

The DIT, depicted in Fig. \ref{transformation}(a), is a very simple transformation which embeds a hidden spin $h$ between 
two interacting visible spins  as follows~\cite{Fisher_Transformations_1959, Wegner_Duality_1971},
\begin{eqnarray}\label{DIT}
e^{ J \sigma_1 \sigma_2} = \Delta \sum_{h = \pm1} \exp \left[ W (\sigma_1  + {\rm sgn} (J)\sigma_2)  h\right],
\end{eqnarray}
where $\Delta$ is the normalization factor. The new interaction $W$ is given as 
\begin{eqnarray}\label{decoration}
W = {\rm arc}\cosh \left(e^{2|J|}\right)/2.
\end{eqnarray}
%\txtrout{where ch denote the hyperbolic cosine function. Other hyperbolic functions, i.e., sinh and tanh, are abbreviated as sh and th as well.}
Since the DIT can be carried out for any $J$, an arbitrary Ising model with two-spin interactions including random spin-glass, frustrated system, and fully-connected models can be mapped into an equivalent RBM. 
Application of such transformation technique allows one to obtain the exact solution for a model on, for instance, two-spin interacting Ising model on a bond-decorated lattice that can be transformed into the soluble model on an undecorated lattice~\cite{Syozi_Statistics_1951, Domb_Green_textbook, Fisher_Transformations_1959, Antonosyan_Exactly_2009, Rojas_Generalized_2009, Lisnyi_Exact_2014}.

The other technique, known as the STT, embeds a hidden spin $h$ into three visible spins with both two- and three-spin interactions as is illustrated in Fig.~\ref{transformation}(b)~\cite{Wegner_Duality_1971, Wu_Eight_1990, Lin_General_1990,  Lu_Soluble_2005, Computation_Caravelli_2018}.
Expressed in the form of the Boltzmann weight, this can be written as~\cite{Lin_Rigorous_1991}
\begin{eqnarray}\label{STT}
\exp\left[M\sigma_1 \sigma_2 \sigma_3 + J_1 \sigma_2 \sigma_3 + J_2 \sigma_3 \sigma_1 + J_3 \sigma_1 \sigma_2\right] \nonumber\\
\ \ \ = \Delta \sum_{h = \pm1}\exp \left[ \sum_{i=1}^3(W_i h + a_i) \sigma_i + b h\right],
\end{eqnarray}
where $M$ and $J_i$ are the amplitudes of three- and two-spin interactions, respectively.
The interaction between the visible and the hidden spins in the extended space is denoted by $W_i$, and the local fields are denoted by  $a_i$ and $b$, respectively.
It can be shown that Eq.~(\ref{STT}) amounts to eight nonequivalent conditions which yields the solutions as
\begin{widetext}
\begin{eqnarray}
&&\exp(4\chi_i a_i) =  \frac{\sinh(2(|J_i| + M))}{\sinh(2(|J_i| - M))},\\ 
&&\cosh(2W_i) = \frac{e^{2|J_i|} \cosh\left(2(|J_j| \!+\! |J_k|)\right) - e^{-2|J_i|}\cosh\left(2(|J_j| \!-\! |J_k|)\right)}{[2\cosh(4|J_i|) - 2\cosh(4M)]^{1/2}}, \label{STT_W}\\
&&\sinh(2b) = \frac{-\sinh(2\chi_i W_i) \sinh(4M)}{[(\cosh(4J_j) - \cosh(4M)) (\cosh(4J_k) - \cosh(4M)) ]^{1/2}}, \\ \nonumber
\label{STT_b}
\end{eqnarray}
\end{widetext}
where $\chi_i = {\rm sgn}(J_i)$ is the sign of the two-spin interaction.
The subscripts in Eqs. (\ref{STT_W}) and (\ref{STT_b}), i.e., $i, j,$ and $k$, must be chosen such that none of them are identical to each other.
Importantly, the STT is valid under the following conditions%~\txtr{\cite{comment1}}
~\cite{comment1}
\begin{eqnarray}
|M|<|J_i|&\ & (i = 1,2,3), \label{STTCondEqn1} \\
\text{sgn}(J_1J_2J_3) &=& 1. \label{STTCondEqn2}
\end{eqnarray}

The notion of the DIT and STT can be generalized to include many-spin interactions. 
A system in the original space with both four- and two-spin interactions, for instance, can be mapped to a model with three-spin interactions that involves a single hidden spin.
This mapping, used to obtain the solution of the zero-field eight-vertex model, is referred to as the star-square transformation~\cite{Baxter_Eight_1971, Baxter_Partition_1972, Wegner_Duality_1971, Streca_Strong_2018}. 
While the transformation from the extended space into the original space, known as ``the star-polygon transformation,'' is achieved by tracing out the hidden spins and is in general tractable~\cite{Rojas_Generalized_2009, Streca_Generalized_2010},  its inverse mapping exists in very limited cases. 
%\txtr{
We note in passing that the DIT and STT have been extended to models with local quantum degrees of freedom such as Heisenberg spins and itinerant electrons~\cite{Canova08, Strecka08, Strecka09}.
%}

\subsection{Generalized Ising model as Boltzmann Machine}
Here, we show that the generalized Ising model, which consists of many-spin interactions, can be mapped to an equivalent model with two-spin interactions and local fields.
First, let us consider some bare many-spin interaction within the model. The Boltzmann factor is given as
\begin{eqnarray}
\pi_{|\mathcal{C}|}(\sset_{\cal C};M) := \exp\left(M\prod_{j \in {\cal C}} \sigma_j\right),
\end{eqnarray}
where $M$ is the amplitude of the interaction.
Here, the set of sites of the visible spins involved, dubbed as the ``cell'' in the following, is denoted as ${\cal C}$.
Correspondingly, a visible spin configuration of the cell is denoted by $\sset_{\cal C} \in {\cal S_{\cal C}}$, where ${\cal S}_{\cal C}$ is the set of all possible configurations of the visible spins included in $\cal C$. 
The subscript of the Boltzmann factor, $|{\cal C}|$, is the number of the visible spins included in the cell.
Note that arbitrary hidden and deep spins can be uniquely labeled due to the construction procedure we introduce in the following.
It is straightforward to see that the Boltzmann weight of the original system can be written as
\begin{eqnarray}\label{IntProductEqn}
\pi(\sset) \!=\! \prod_{\cal C} \pi_{|\cal C|}(\sset_{\cal C}; M_{\cal C}) \!=\! \prod_{\cal C} \left( \sum_{\hset_{\cal C}, \dset_{\cal C}} \widetilde{\pi}_{\cal C}(\sset_{\cal C}, \hset_{\cal C}, \dset_{\cal C}) \right),
\end{eqnarray}
where $\widetilde{\pi}_{\cal C}(\sset_{\cal C}, \hset_{\cal C}, \dset_{\cal C})$ is the weight of the RBM obtained by transformation. 
Note that although a visible spin may be included in multiple cells, the auxiliary spins are uniquely allocated to their corresponding cells by construction. 
A single realization of hidden and deep spins labeled by ${\cal C}$ is denoted as $\hset_{\cal C} \in {\cal H_{\cal C}}$ and $\dset_{\cal C} \in {\cal D_{\cal C}}$, where ${\cal H}_{\cal C}$ and ${\cal D_{\cal C}}$ are the sets of all possible configurations of the hidden and deep spins included in $\cal C$, respectively.

As we see from Eq. (\ref{IntProductEqn}), the Boltzmann factor of any model can be decomposed into products over cells and the independence of hidden spins holds. 
For simplicity, we focus on bare $p$-spin interactions in the following.

\subsubsection{Three-spin interaction as RBM}\label{3spinAsRBMSection}
\begin{figure}[t]
\hspace{-1.5cm}
  \begin{minipage}{0.85\hsize}  
     \resizebox{!}{0.8\hsize}{\includegraphics{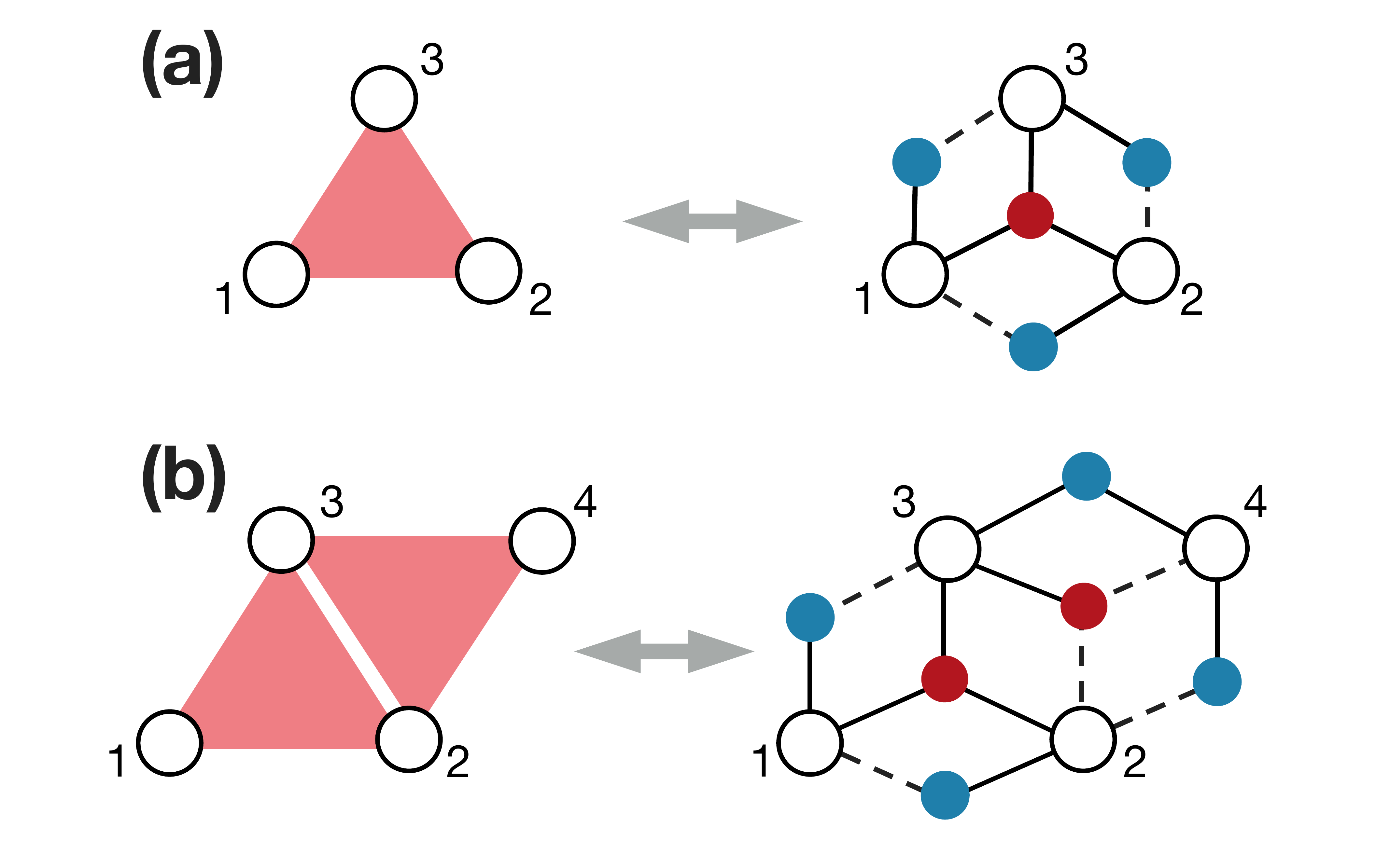}}
     \resizebox{1.1\hsize}{!}{\includegraphics{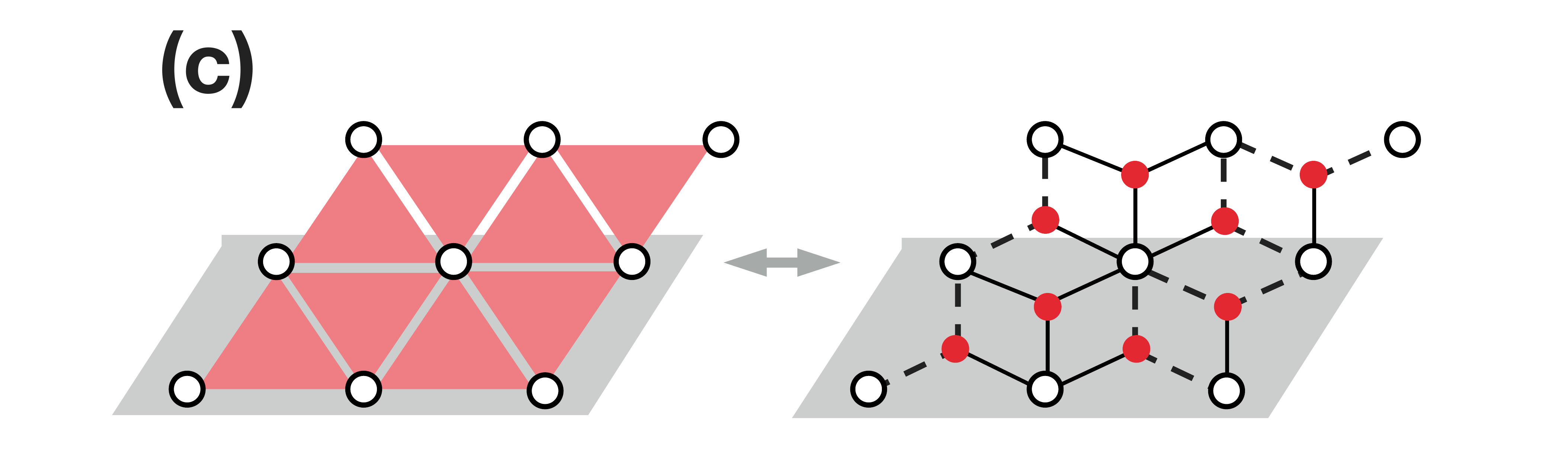}}
  \end{minipage}
\caption{\label{3bodiesAsRBM}(Color Online) (a) Transforming a pure three-spin interaction into an RBM. 
The red filled circles in the right-hand side are generated by the STT and the blue ones by the DIT. The black solid and dotted lines denote the positive and negative two-spin interactions, respectively.  The amplitudes of the virtual two-spin interactions are taken as $J_i>0$ in the figure. (b) Transforming a couple of three-spin interactions. The signs of the virtual two-spin interactions are modified so that the number of the hidden spins is reduced.  (c) Transforming the Baxter-Wu model into an RBM.}
\end{figure}

First, we discuss the three-spin interaction, which is the simplest possible case.
The STT cannot be applied straightforwardly to a bare three-spin interaction due to the conditions given in Eqs. (\ref{STTCondEqn1}) and (\ref{STTCondEqn2}).
To avoid this problem, we introduce the ``virtual two-spin interactions'' that cancel each other out as follows,
\begin{eqnarray}\label{3spinSTT}
\pi_3(\sset;M) &=& \exp(M \sigma_1 \sigma_2 \sigma_3) \nonumber \\
&=& \exp\left[M\sigma_1 \sigma_2 \sigma_3 + J_1\sigma_2\sigma_3 + J_2 \sigma_3\sigma_1 + J_3 \sigma_1 \sigma_2\right] \nonumber\\
&\times& \exp\left[- (J_1 \sigma_2 \sigma_3 + J_2 \sigma_3 \sigma_1 + J_3\sigma_1 \sigma_2)\right].
\end{eqnarray}
As is depicted in Fig. \ref{3bodiesAsRBM}(a), this can be mapped into the RBM by applying Eq. (\ref{STT}) to the first and subsequently Eq. (\ref{DIT}) to the second factors.
Note that the amplitudes of the virtual two-spin interaction, $J_i$, can be taken arbitrarily as long as Eqs. (\ref{STTCondEqn1}) and (\ref{STTCondEqn2}) are satisfied.

Next, let us consider two sets of interacting spins as denoted in Fig. \ref{3bodiesAsRBM}(b). 
Although the naive application of Eq. (\ref{3spinSTT}) yields eight hidden spins, two of them on the shared edges can be eliminated by modifying the signs of virtual two-spin interactions.
For instance, by considering the virtual two-spin interactions with a homogeneous amplitude, we obtain
\begin{eqnarray}
&&\pi_3(\sset_{{\cal C}_1}; M) \pi_3(\sset_{{\cal C}_2}; M) \nonumber \\
 &=& \exp(M \sigma_1 \sigma_2 \sigma_3) \exp(M \sigma_2 \sigma_3 \sigma_4) 
\nonumber \\
&=& \exp\left[M\sigma_1 \sigma_2 \sigma_3 + J\sigma_2\sigma_3 + J \sigma_3\sigma_1 + J \sigma_1 \sigma_2\right] \nonumber\\
&&\ \ \times \exp\left[M\sigma_2 \sigma_3 \sigma_4 + J\sigma_3\sigma_4 - J \sigma_4\sigma_2 - J \sigma_2 \sigma_3\right] \nonumber \\
&&\ \ \times \exp\left[- (J \sigma_3 \sigma_1 + J\sigma_1 \sigma_2  + J\sigma_3 \sigma_4 - J\sigma_4 \sigma_2)\right],
\end{eqnarray}
from which six hidden spins emerge.

Furthermore, a system with pure three-spin interaction on the triangular lattice, known to be exactly soluble and dubbed as the
 Baxter-Wu model~\cite{Baxter_Exact_1973}, can also be mapped into the RBM merely without hidden spins generated by the DIT. 
 In other words, we may choose the signs of the virtual two-spin interactions on the triangles carefully so that the sum at each edge would cancel out, resulting in a reduced number of the hidden spins.
Still, there are exponential number of ways to represent this model by tuning the amplitudes and the signs of the virtual two-spin interactions.
Shown in Fig. \ref{3bodiesAsRBM}(c) is the mapping with 4-fold periodicity along the $x-$axis.

%----------------------------------------------------%
%----------------------------------------------------%

\subsubsection{Four-spin interaction as DBM}
Next, we show that four-spin interaction can be expressed  by introducing the second hidden layer, or the ``deep'' layer.
The illustration of the two-step transformation is shown in Fig. \ref{NbodiesAsDBM} (a).
In the first step, we interpret the product of Ising variable as a single new binary variable and apply the DIT as
\begin{eqnarray}
\pi_4(\sset; M^{(0)}) &=& \exp \left[M^{(0)} \sigma_1 \sigma_2 \sigma_3 \sigma_4 \right]\nonumber \\
 &=& \Delta \sum_{d = \pm1}\exp\left[M^{(1)}(\sigma_1 \sigma_2 + \sigma_3 \sigma_4)d \right] \nonumber\\
 &=& \Delta \sum_{d = \pm1} \pi_3 \left(\sigma_1, \sigma_2, d; M^{(1)}\right) \pi_3 \left(\sigma_3, \sigma_4, d; M^{(1)}\right), \nonumber \\
\ 
\end{eqnarray}
where $\pi_4(\sset; M^{(0)})$ is the Boltzmann factor for four-spin interaction with the amplitude $M^{(0)}$. 
The interaction amplitudes $M^{(0)}$ and $M^{(1)}$ are related by Eq. (\ref{decoration}) as 
\begin{eqnarray}
M^{(1)} = {\rm arc}\ \cosh\left(e^{2|M^{(0)}|}\right)/2.
\end{eqnarray}

Next, we use Eq. (\ref{3spinSTT}), the result for the three-spin interaction, under homogeneous virtual two-spin interactions $J  > |M^{(1)}|$ for simplicity.
Here, we obtain the expression as
\begin{flalign}
\pi_4(\sset; M^{(0)}) &= \Delta \sum_{d}\sum_{h_a, h_b}\sum_{h_1, ... h_6}
\widetilde{\pi}_{2}^{\rm STT}(\sset, \hset,d) \widetilde{\pi}_{2}^{\rm DIT}(\sset, \hset,d)  \nonumber \\
&\hspace{3.5cm} \times \widetilde{\pi}_{1}(\sset, \hset, d), 
\end{flalign}
where
\begin{flalign}
\widetilde{\pi}_{2}^{\rm STT}(\sset, \hset,d) &=\exp[W(\sigma_1 + \sigma_2 + d)h_a + W(\sigma_3 + \sigma_4 + d)h_b], \nonumber \\ 
\widetilde{\pi}_{2}^{\rm DIT}(\sset, \hset, d)&=\exp\left[\sum_{i=1}^4W'(\sigma_i - d)h_i \right.\nonumber \\
&\ \ \ \ \ \ \ \  + W'(\sigma_1 - \sigma_2)h_5 + W'(\sigma_3 -\sigma_4)h_6 \Bigg] \nonumber, \\
\widetilde{\pi}_{1}(\sset, \hset,d) &=\exp\left[a \sum_i\sigma_i + a_d d + b (h_a + h_b) \right].
\end{flalign}
Here, $\widetilde{\pi}_{2}^{\rm STT({\rm DIT})}(\sset, \hset,d)$ is the Boltzmann factor for the two-spin interaction in the extended space obtained by the STT (DIT), and $\widetilde{\pi}_1(\sset, \hset, d)$ is that for the local external fields. The interaction $W$ $(W')$ is the interlayer coupling introduced by the STT (DIT). 
The magnetic field for the visible, deep, and hidden spins are denoted by $a, a_d, $ and $b$, respectively. 
The equivalence of amplitudes of the three-spin interactions, $M^{(1)}$, leads to $a_d = 2a$. 
Also, the application of DIT to the virtual two-spin interaction yields
$W' = {\rm arc}\cosh(e^{2J})/2.$
Note that the other parameters  are obtained by substituting $M = M^{(1)}$ and homogeneous $J_i = J$ into Eq. (\ref{STT}) as 
\begin{eqnarray}
\exp(4a) &=&  \frac{\sinh(2(J + M^{(1)}))}{\sinh(2(J - M^{(1)}))},\\ 
\cosh(2W) &=& \frac{e^{2J} \cosh(4J) - e^{-2J}}{[2\cosh(4J) - 2\cosh(4M^{(1)})]^{1/2}}, \\
\sinh(2b) &=& \frac{-\sinh(2W) \sinh(4M^{(1)})}{|\cosh(4J) - \cosh(4M^{(1)}))| }.
\end{eqnarray}

Another way to transform four-spin interaction is to apply the star-square transformation. 
Although it also requires a single deep spin, the architecture of the hidden spins would be symmetric and hence different from the aforementioned transformation.
To keep the number of auxiliary spins minimum, we will not use the star-square transformation in the following.

Note that $N_h/N_d$, or the ratio of the number of the hidden spins to that of the visible spins, may be reduced in a larger system as well as in Sec. \ref{3spinAsRBMSection}; we may cancel out the virtual two-spin interactions by modifying their signs.

\subsubsection{$p$-spin interaction as DBM}\label{pspinSection}
The discussion for four-spin interaction can be extended to $p$-spin interaction.
Assume that Ising interactions up to $(p-1)$-spin terms can be mapped into a BM. 
The DIT splits the original model into two $(p/2 + 1)$-spin terms for even $p$ and $((p+1)/2 + 1)$- and $((p-1)/2 + 1)$-spin terms for odd $p$.
The mapping exists for $p=3,4$ as we have shown in the previous sections, hence for arbitrary $p>4$ as well.
For further discussion and explicit expression in the form of BM, see Appendix \ref{pspin2DBMSection}.

\begin{figure}[t]
\begin{center}
\begin{tabular}{c}
  \begin{minipage}{0.97\hsize}
    \begin{center}
     \resizebox{0.95\hsize}{!}{\includegraphics{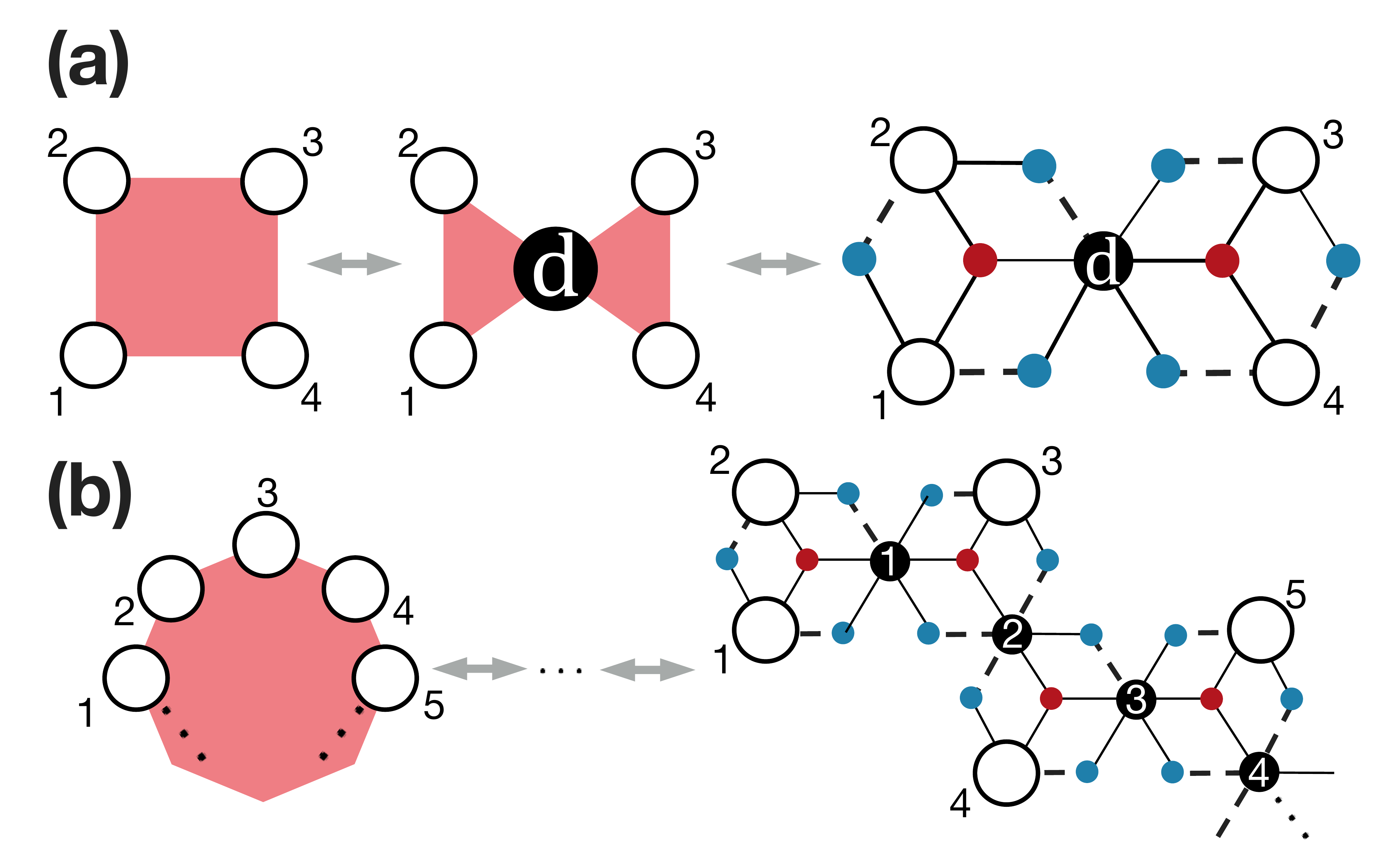}}

   \end{center}
  \end{minipage}
\end{tabular}
\end{center}
\caption{\label{NbodiesAsDBM}(Color Online) (a) Transforming four-spin interaction into a DBM. The black filled circle is the spin in the deep layer, and the other notation follows that of Fig. \ref{3bodiesAsRBM}. (b) Transforming $p$-spin interaction into a DBM. The architecture of the BM is determined from the number of the spins involved in the interaction, but is irrelevant to the actual spatial dimension.
%\txtb{[The font of d in the figure has been changed due to illustrator...]}
%\txtb{[We should put numbers on hidden neurons as Fig. 8?]}
%\txtb{[NY: I have added numbers on hidden neurons newly in v2.2 as Akagi-san has suggested.]}
}
\end{figure}

The number of the hidden and deep spins can be computed as well. We denote the numbers of hidden spins in the first layer generated by the STT, DIT, and that of the deep spins in the second hidden layer as $n_h^{\rm STT}, n_h^{\rm DIT},$ and  $n_d$, which are shown in blue, red, and black filled circles in Fig. \ref{NbodiesAsDBM}, respectively.
It is straightforward to show that the following relations are satisfied,
\begin{eqnarray}
n_h^{\rm STT} = 3(p-2),\quad n_h^{\rm DIT} = p - 2,\quad  n_d = p - 3.
\end{eqnarray}
It is noteworthy from the perspective of the numerical cost that the number of hidden and deep spins increases only linearly with respect to $p$. The computational order remains to be the same for any update scheme.

%----------------------------------------------------%
%----------------------------------------------------%

\section{Monte Carlo sampling on Boltzmann machine}\label{MCSection}
In the following, we utilize the BM obtained by the transformation to classical MC sampling.
The block Gibbs sampling, which is frequently used in the machine learning community, turns out to lack efficiency in terms of autocorrelation although the numerical cost per a single MC step is low~\cite{Goodfellow_DeepLearning_2016, Huang_Accelerated_2017}. 
We alternatively apply the Swendsen-Wang algorithm to the extended space and demonstrate the speed up compared to the SSF on the original space.
First, we take the square-lattice Ising model for a simplified description of our scheme, and then proceed to show the results in the model with ferromagnetic two-spin interactions and alternating three-spin interactions on the Kagom\'e lattice, which is one of the the most comprehensible models that includes the many-spin interaction and also suffers from the slowing down at the critical temperature.

\subsection{Ising model on square lattice}
We consider a Hamiltonian with two-spin interactions on a square lattice,
\begin{eqnarray}\label{2DIsing}
H = -\sum_{\langle i,j \rangle}\sigma_i \sigma_j,
\end{eqnarray}
which shows the ferro-paramagnetic transition at $T_{\rm c} = 2/\ln(\sqrt{2} + 1)$ as is widely known in statistical physics~\cite{Onsager_Ising_1944}.
The study by the SSF algorithm, i.e., the Glauber dynamics in a broader sense,  suffers from severe critical slowing down~\cite{Glauber_Timedependent_1963, Ito_Nonequilibrium_1993}.
The application of global updates, e.g., cluster updates~\cite{Swendsen_Nonuniversal_1987, Wolff_Collective_1989}, is one of the solutions under some circumstances.

Note that Eq. (\ref{2DIsing}) itself can be regarded as the RBM since the square lattice is bipartite. 
Since the block Gibbs sampling method, which fix the configuration in a sublattice to enable independent sampling from another, is not beneficial as is qualitatively discussed in Appendix \ref{GibbsSection}, 
we alternatively apply the well-known Swendsen-Wang algorithm to take advantage of the expression.
Although there is substantially no further gain for two-spin interacting model, in the following subsection, we demonstrate the speed up in terms of the autocorrelation time. As is shown in Sec. \ref{transformSection}, the number of the hidden and deep spins increases linearly with the system size, and therefore the computational order of a single MC step remains to be $O(N)$ in a model with short-range interactions, which is also the case in the current work.

\subsection{Generalized Ising model on Kagom\'e lattice}
Here, we consider a model with ferromagnetic two-spin interactions and also three-spin interactions on a Kagom\'e lattice. 
Let $E$ be the set of edges and $\up (\down)$ be an upward (downward) triangle on the lattice.
The Hamiltonian is defined as
\begin{eqnarray}\label{kagomeCheckerHam}
-\beta H(\sset) = \beta\sum_{\langle i, i'\rangle \in E}\sigma_i \sigma_{i'} + \sum_{\up}M_{\up}\tau_{\up} + \sum_{\down}M_{\down}\tau_{\down},
\end{eqnarray}
where $\tau_{\up} = \prod_{i \in \up}\sigma_i$ and $\tau_{\down} = \prod_{i \in \down}\sigma_i$ is the product of the spin variables, $M_{\up}/\beta\ (M_{\down}/\beta)$ denotes the amplitude of the three-spin interactions for upward (downward) triangles, and $\langle i, i'\rangle$  is the edge connecting sites $i$ and $i'$.
The symmetry that combines the spin inversion and the mirror inversion is present when $M_{\up} + M_{\down} = 0$, and the model exhibit second-order transition at finite temperature for finite $M_{\up}$ [See Appendix \ref{PartitionSection} for discussion when only three-spin interactions are present.]. 
This can be understood from the correspondence between the present model and the antiferromagnetic Ising model with a uniform external field on the honeycomb lattice~\cite{Wu_Critical_1989}.

%----------------------------------------------------%
\begin{figure}[tb]
\begin{center}
\begin{tabular}{c}
  \begin{minipage}{0.98\hsize}
    \begin{center}
     \resizebox{0.99\hsize}{!}{\includegraphics{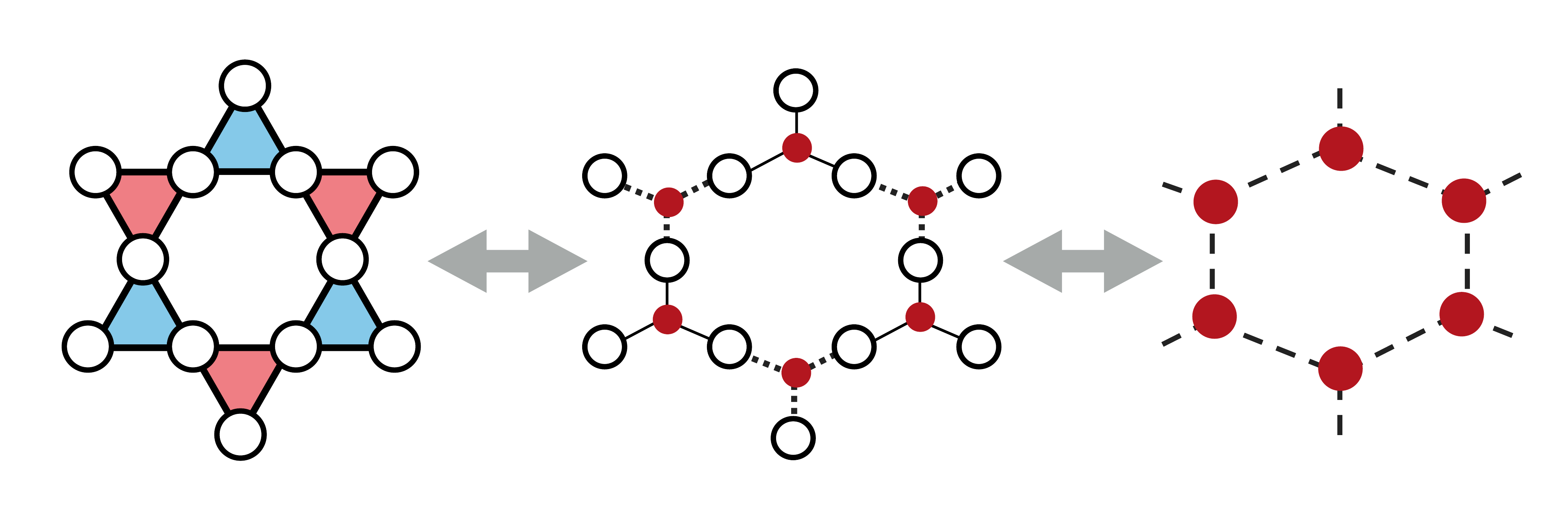}}

   \end{center}
  \end{minipage}
\end{tabular}
\end{center}
\caption{\label{kagomeCheckerAsRBM} (Color Online) Graphical understanding of the transformation for the spin model defined by Eq. (\ref{kagomeCheckerHam}) into RBM and hidden-spin-only model. 
The original space is defined on the Kagom\'e lattice, in which the triangles are colored to denote the signs of three-spin interactions. 
The white open and red filled circles correspond to the visible and hidden spins, respectively. 
Also, the signs of the two-spin interactions are represented as solid and dotted lines for positive and negative, respectively.}
\end{figure}
%----------------------------------------------------%

In the following, we assume $M_{\up} = -M_{\down} = M>0$.
The Boltzmann weight is transformed as
\begin{eqnarray}
\pi(\sset) &=& \exp\left[\beta\sum_{\langle i, i'\rangle \in E}\sigma_i \sigma_{i'} + \sum_{\up}M_{\up}\tau_{\up}  
+ \sum_{\down}M_{\down}\tau_{\down} \right] \\
&=& \Delta \sum_{\hset}\exp\left[ \sum_{\up} W_{\up} h_{\up} \small\sum_{i\in \up}\sigma_i  + \sum_{\down} W_{\down} h_{\down}\sum_{i\in \down}\sigma_i  \right. \nonumber\\
&&\ \ \ \ \ \ \ \ \ \ \ \ \ \ \ \ \ \ \ \ \ \ \ \ \ \ \ \ \left. + b\sum_{\up, \down} (h_{\up} + h_{\down})\right] \nonumber\\
&=& \sum_h \widetilde{\pi}(\sset, \hset),
\end{eqnarray}
where the parameters in the extended model are obtained by substituting $M_{\up(\down)}$ and $\beta$ into the STT, or Eq. (\ref{STT}). 
Note that the external fields on the visible spins is absent due to the cancellation caused by the alternating signs of the interactions.
As is graphically described in Fig. \ref{kagomeCheckerAsRBM}, one hidden spin is embedded per triangle and denoted as $h_{\up}$ or $h_{\down}$.
The explicit expressions for the parameters can be read off from
\begin{eqnarray}
\cosh(2W_{\up(\down)}) &=& \frac{e^{2\beta} \cosh(4\beta) - e^{-2\beta}}{[2\cosh(4\beta) - 2\cosh(4M)]^{1/2}}, \\
\sinh(2b) &=& \frac{-\sinh(2W) \sinh(4M)}{|(\cosh(4\beta) - \cosh(4M))|},
\end{eqnarray}
where $W_{\up} = -W_{\down} = W>0$ and the signs of the parameters satisfy ${\rm sgn}(W_{\up(\down)}) = {\rm sgn}(M_{\up(\down)})$.
Owing to the cancellation of the magnetic field on visible spins, we obtain a simple expression by tracing out the visible spins as
\begin{eqnarray}
\widetilde{\widetilde{\pi}}(\hset) &=& \sum_{\sset} \widetilde{\pi}(\sset, \hset)\nonumber\\
&=& \Delta' \exp \left(-W_h\sum_{\langle j, j' \rangle \in \overline{E}} h_j h_{j'} + b\sum_{j \in \overline{V}} h_j\right), \label{honeycombHiddenHam}
\end{eqnarray}
where $\Delta'$ is another renormalization factor, $\overline{E}$ and $\overline{V}$ are the sets of edges and vertices in the honeycomb lattice as is shown in the rightmost panel of Fig. \ref{kagomeCheckerAsRBM}.
Note that the interaction is antiferromagnetic, reflecting the alternating signs of the three-spin interactions in the original space.
The amplitude of the two-spin interaction is obtained by the DIT as
\begin{eqnarray}
W_h = {\rm arc}\cosh \left(e^{2W}\right)/2.
\end{eqnarray}
Although the model defined by Eq. (\ref{honeycombHiddenHam}) is not soluble at $b\neq0$, 
an approximate solution of the transition point can be obtained by imposing some assumption after mapping the original model to an eight-vertex model~\cite{Wu_Critical_1989}. 
This turns out to be  in a fairly good precision but not exact, and hence we determine the transition point from the finite size scaling of the Binder ratio~\cite{Binder_Critical_1981, Tanaka_Quantum_2017}. 

 %----------------------------------------------------%
\begin{figure}[b]
\begin{center}
\begin{tabular}{c}
  \begin{minipage}{0.9\hsize}
    \begin{center}
     \resizebox{0.99\hsize}{!}{\includegraphics{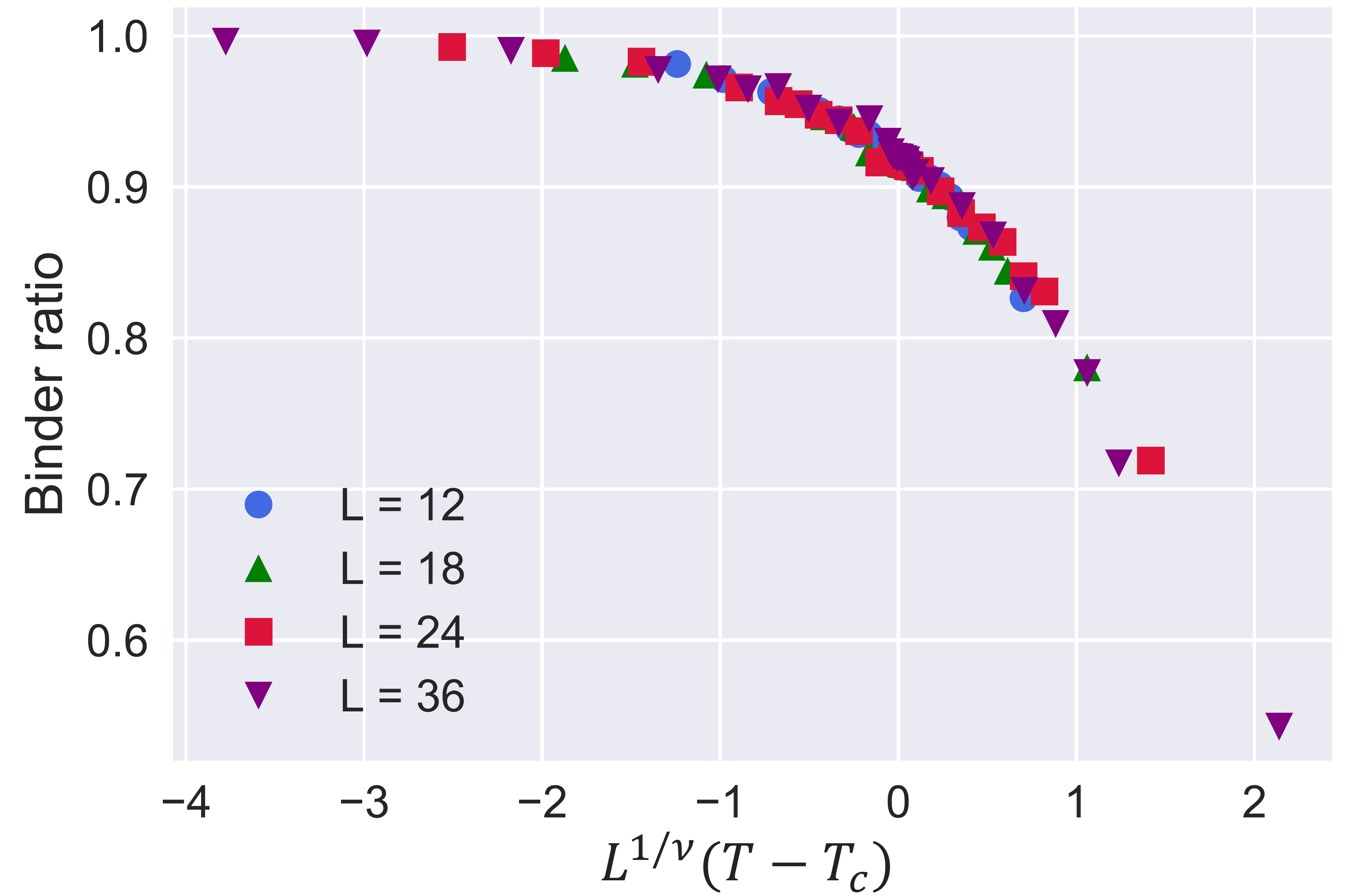}}
   \end{center}
  \end{minipage}
\end{tabular}
\end{center}
\caption{\label{BinderFig}(Color Online) Data collapse of the Binder ratio for the Ising model with two-spin and alternating-sign three-spin interaction on the Kagom\'e lattice. The critical temperature and the critical exponent for the correlation length are obtained as $T_{\rm c} \sim 2.141$ and $\nu \sim 0.99$, respectively. 
The magnitude of two-spin interactions is set to unity while that of  three-spin interactions is $M/\beta_{\rm c}$ = 0.1. The blue circles, green upward triangles, red squares, and purple downward triangles denote the data for the linear system sizes $L = 12, 18, 24,$ and $36$, respectively.}
\end{figure}
%----------------------------------------------------%

 %----------------------------------------------------%
\begin{figure}[t]
\hspace{-0.5cm}
  \begin{minipage}{1.05\hsize}
    \begin{center}
     \resizebox{0.98\hsize}{!}{\includegraphics{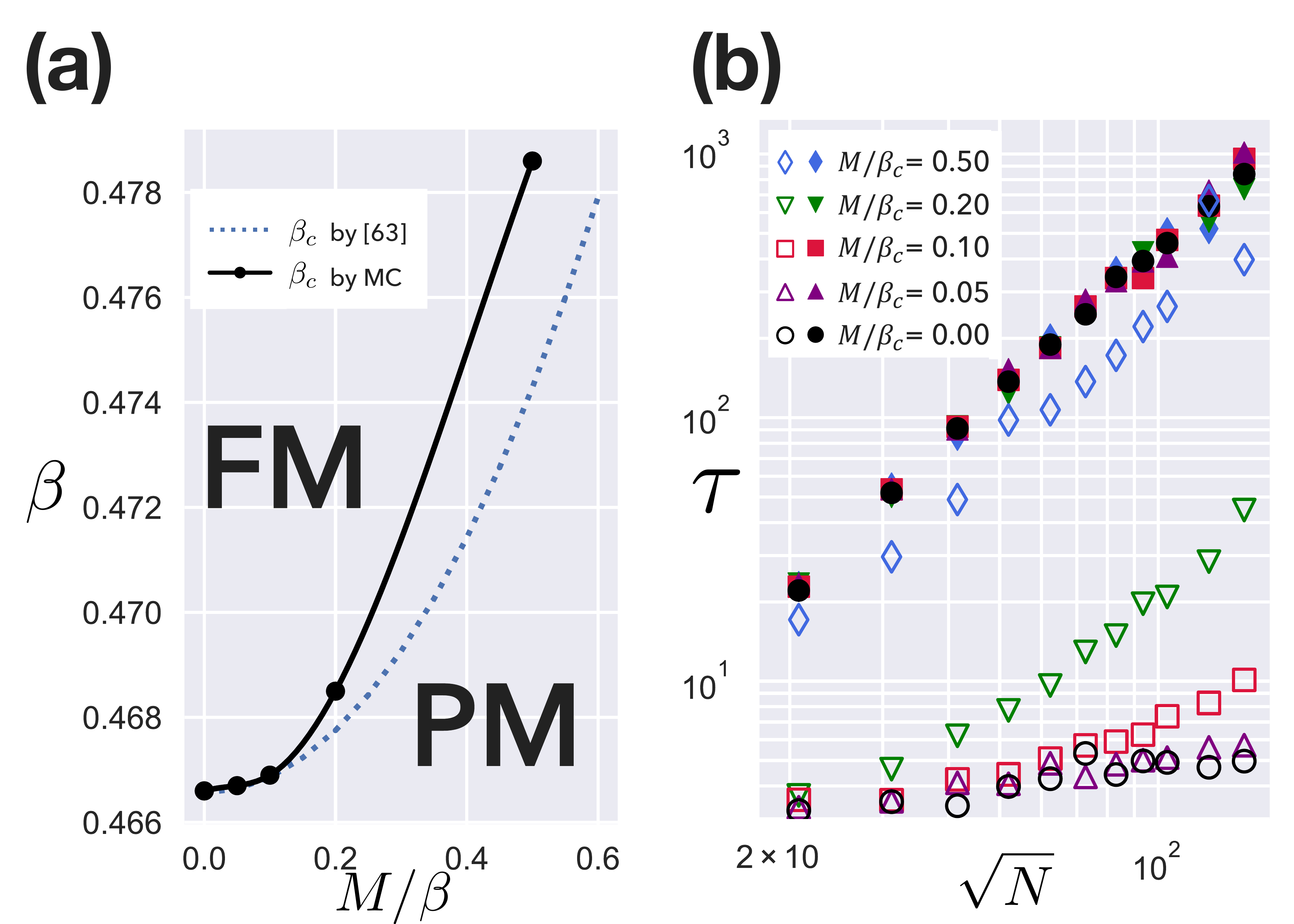}}
   \end{center}
  \end{minipage}
\caption{\label{AutoCorr_kagomeChecker}(Color Online) (a) The autocorrelation time of the magnetization measured in units of Monte Carlo step. The black circle, purple upward triangle, red rectangle, green downward triangle, and blue diamond markers denote the magnitude of three-spin interactions to be $0, 0.05, 0.1, 0.2, $ and 0.5, respectively. 
The filled (unfilled) markers represent the results by the single-spin flip in the original space (cluster update in the extended space). (b) The phase diagram of the model defined in Eq. (\ref{kagomeCheckerHam}). The boundary between the ferromagnetic (FM) and the paramagnetic (PM) phases are given here. The blue dashed line is calculated from the approximate solution in Ref.~\cite{Wu_Critical_1989}, and the black dots are given by the finite-size scaling of the Binder ratio. The numerically estimated inverse critical temperature at $M = 0$ approaches $\beta_{\rm c} = \ln(3 + 2\sqrt{3})/4 \sim 0.4666$, which can be obtained from the exact solution~\cite{Syozi_Statistics_1951}.}
\end{figure}
%----------------------------------------------------%

As was introduced by Binder, the renormalization group theory leads us to assume the scaling law for the Binder ratio, 
\begin{eqnarray}
g:=\frac{1}{2}\left(3 - \frac{\langle m^4 \rangle}{\langle m^2 \rangle^2}\right),
\end{eqnarray}
where $m$ is the magnetization per site and $\langle \cdots \rangle$ denotes the thermal average.
The scaling of this quantity in the vicinity of the critical temperature, $T_{\rm c}$, is given as follows,
\begin{eqnarray}
g \sim  F\left(L^{1/\nu}(T-T_{\rm c})\right),
\end{eqnarray}
where $L$ is the linear system size, $\nu$ is the critical exponent for the correlation length, and $F$ is an appropriate polynomial function, which has been taken to be cubic in this work.
Shown in Fig. \ref{BinderFig} are the results for ferromagnetic two-spin and alternating-sign three-spin interacting model on the Kagom\'e lattice with $M/\beta = 0.1$. 
From the data collapse in Fig. \ref{BinderFig}, we see that the scaling analysis is valid. 
The critical exponent $\nu$ is confirmed for numerous $M/\beta$ to be in good agreement with $\nu = 1$, suggesting that the transition falls into the two-dimensional Ising universality class~\cite{Fisher_Rigorous_1969}.
The simultaneously estimated quantity, i.e., the critical temperature $T_{\rm c}$, is summarized in Fig. \ref{AutoCorr_kagomeChecker}(a) together with the approximate solution given by Ref. \cite{Wu_Critical_1989}.

At the critical temperature, we use the cluster algorithm in the extended space. The detailed description is given in Appendix \ref{MagClusterSection}.
Our main result is summarized in Fig. \ref{AutoCorr_kagomeChecker}(b), in which we compare the autocorrelation time $\tau$ of the magnetization measured in units of Monte Carlo steps per site for the whole system. [See Appendix \ref{ObservationSection} for the calculation of physical quantities in the extended space. ]
The magnetization at $t$-th Monte Carlo step is calculated from the spin configurations $\sset(t) = (\sigma_1(t), \sigma_2(t), ..., \sigma_{N_v}(t))$ with $N_v$ being the number of visible spins as
\begin{eqnarray}
m(t) = \sum_{i=1}^{N_v} \frac{\sigma_i(t)}{N_v},
\end{eqnarray}
The estimation of $\tau$ is done by evaluating the decay of the equilibrium autocovariance \cite{Evertz_Loop_2003},
\begin{eqnarray}
A(t) = \frac{\langle |m(t_0 + t)m(t_0)|\rangle - \langle |m(t_0)| \rangle^2}{\langle|m(t_0)|^2\rangle - \langle|m(t_0)|\rangle^2} = A_0 e^{-t/\tau},
\end{eqnarray}
where $\langle \cdots \rangle$ denotes the average over $t_0$, namely the MC steps.
The critical slowing down is constantly observed in the SSF performed on the original space, while the application of the cluster update to the BM significantly improves the situation. We observe that the dynamical exponent $z$, which is the slope of data in Fig. \ref{AutoCorr_kagomeChecker}(b), is also reduced, while the possibility that it gradually grows in the larger system sizes cannot be ruled out.

The increase in the autocorrelation time along the three-spin interaction $M/\beta_{\rm c}$ is understood in the following way. 
The virtual magnetic field induced in the hidden spins by the STT is amplified as  $M/\beta_{\rm c}$ is increased,  and thus the virtual magnetization per cluster increases, modifying the flipping probability of each cluster to be unbalanced.
While the cluster is flipped randomly when the magnetic field is absent, finite-valued Zeeman energy results in unbalanced flipping probability due to the detailed balance condition. [See Appendix \ref{MagClusterSection} for further discussion.]
Such a situation prevents the system from exploring the spin configurations efficiently, and thus show a weaker speed up.

\section{Conclusion and Discussion}\label{DiscussionSection}
In the current work, we found an algebraic transformation of the Ising model with many-spin interactions into the Boltzmann machine in which only two-spin interactions and virtual local fields are present.
The decoration-iteration and star-triangle transformations were applied to embed hidden and deep spins, namely the auxiliary degrees of freedom to be traced out.
At the expense of the dimension of the spin space, significant suppression of the critical slowing down is achieved by applying the cluster algorithm to the Boltzmann machine.

Our scheme is also capable of handling  continuous classical spin systems with many-spin interactions.
As in the case with two-spin interactions \cite{Wolff_Collective_1989}, we may consider projecting each variable on some axis. 
Namely, we rewrite a continuous variable ${\bm S}_i$ on site $i$ by a new Ising variable $\sigma_i$ as ${\bm S}_i = \sigma_i |{\bm S}_i \cdot {\bm n}_i| {\bm n}_i + {\bm S}_i^{\perp}$, where ${\bm n}_i$ is the randomly chosen projection axis and ${\bm S}_i^{\perp}$ is orthogonal to ${\bm n}_i$. 
We can now apply our method by regarding the model for $\{\sigma_i\}$ obtained by projection as the generalized Ising model.
The randomness of the projection axis at each Monte Carlo step would assure the ergodicity of the scheme.

Beyond our scope in the current work is the optimal transformation for simulation.
The transformation is non-unique when the virtual interaction is required, 
and we may even consider infinitely strong coupling to express extreme situations such as decoupled or completely aligned pairs of spins.
Non-uniqueness arises also when four-spin interaction is present.
Although we excluded the application of the star-square transformation for clarity, comparison of the 
numerical efficiency between different transformations may be worth investigating. 
Switching into different Boltzmann machines for each step may allow us to explore the free energy landscape more efficiently.

In closing, we would like to note the applicability of the algebraic transformation to wider fields of research.
One interesting direction is undoubtedly the pursuit of equilibrium statistical physics, which includes extending and exploring exactly soluble models and replacing the Swendsen-Wang algorithm with other global updates to tackle frustrated systems.
Another problem lies in the field of computer science; the application to combinatorial optimization problems; 
%\txtb{The decompositions introduced in our paper can be thought of as a finite temperature generalization of the one discussed in Ref.~\cite{Leib_Transmon_2016},} and hence
%\txtr{
our decompositions applicable also in the finite-temperature case may open a new way to introduce ancilla spins required in the experimental implementation of annealing process.
%}

\section*{Acknowledgements}
The authors acknowledge helpful discussions with Seiji Miyashita, Tsuyoshi Okubo, Shu Tanaka, Synge Todo, and Lei Wang. This work was supported by JSPS KAKENHI Grant Nos. JP17K14352. JP18H04220, 18K03445, 18H04478. N. Y. was supported by the  JSPS through Program for Leading Graduate Schools (ALPS) and JSPS fellowship (JSPS KAKENHI Grant No. JP17J00743).

\appendix
 %----------------------------------------------------%
 \begin{figure*}[t]
\begin{center}
\includegraphics[width=1.99\columnwidth]{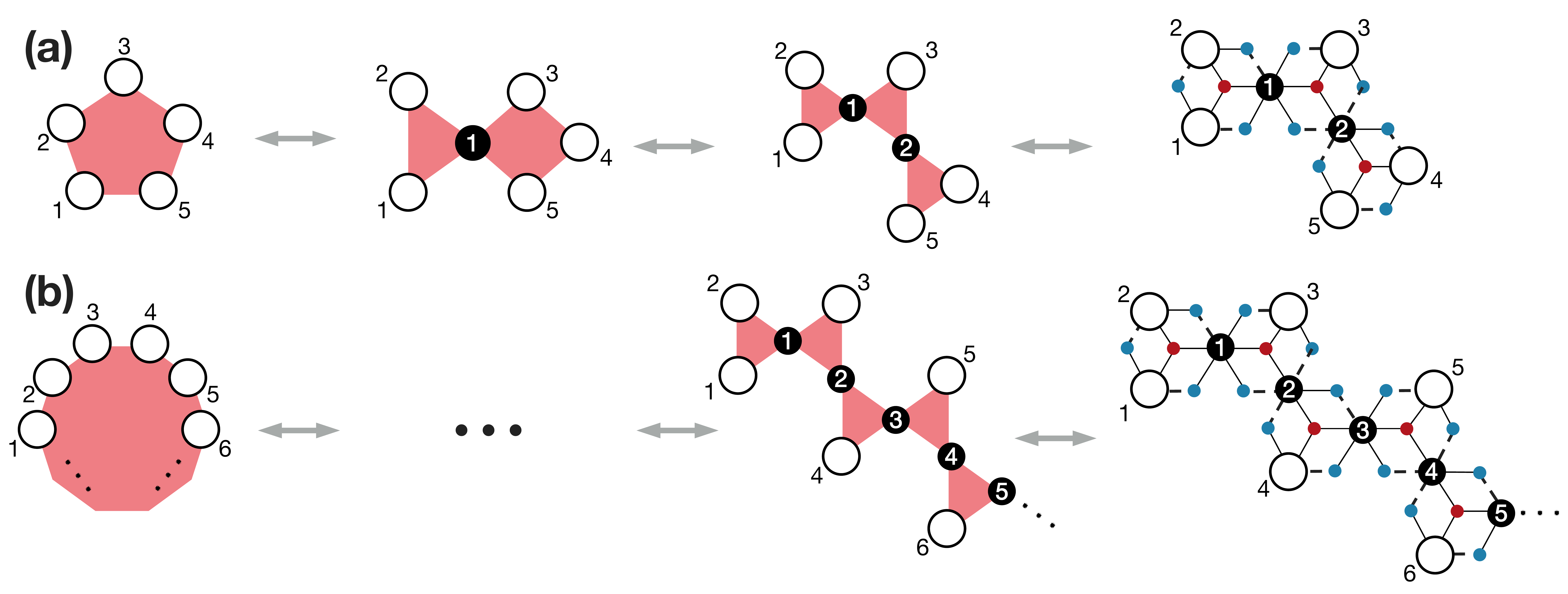}
\caption{\label{NspinTransformationStep} (Color Online) The transformation of (a) five-spin interaction and (b) $p$-spin interaction. First mapped into a system with only three-spin interactions by applying DIT, the transformation technique for the three-spin interaction is used to break up into two-spin interactions with local external fields.}
\end{center}
\end{figure*}
%----------------------------------------------------%
\section{Transformation of $p$-spin interaction}\label{pspin2DBMSection}
In the following, we discuss the transformation of a bare $p$-spin interaction into the BM.
The procedure consists of two steps; embedding the deep spins, and then the hidden spins. 
 A system with a bare $p$-spin interaction is mapped into an equivalent system with three-spin interactions in the former step and subsequently broken into systems with two-spin interactions and local fields in the latter step.
Since the DIT is applied repeatedly, we define
\begin{eqnarray}
M^{(k+1)} &=& {\rm arc}\cosh\left(e^{2|M^{(k)}|}\right)/2, \\
M^{(0)} &=& M,
\end{eqnarray}
where $M$ determines the amplitude of the interaction as $\pi_p(\sset;M)$.
The number of DITs applied, or $k$, is referred to the DIT transformation order.

Let us take four-spin interaction as the starting point. 
As we have shown in the main text, the Boltzmann factor can be expressed by embedding the deep spin as
\begin{eqnarray}
\pi_4(\sset; M^{(0)}) &=& \exp \left[M^{(0)} \sigma_1 \sigma_2 \sigma_3 \sigma_4 \right]\nonumber \\
 &=& \Delta \sum_{d = \pm1} \pi_3 \left(\sigma_1, \sigma_2, d; M^{(1)}\right) \pi_3 \left(\sigma_3, \sigma_4, d; M^{(1)}\right). \nonumber \\
 \ 
\end{eqnarray}
By replacing one of the Ising variables with a product of two, we obtain the expression for five-spin interaction as
\begin{eqnarray}
\pi_5(\sset; M^{(0)}) &=& \exp \left[M^{(0)} \sigma_1 \sigma_2 \sigma_3 \sigma_4 \sigma_5 \right]\nonumber \\
 &=& \Delta \sum_{d_1 = \pm1} \pi_3 \left(\sigma_1, \sigma_2, d_1; M^{(1)}\right) \pi_4 \left(\sigma_3, \sigma_4, \sigma_5, d_1; M^{(1)}\right) \nonumber\\
 &=& \Delta \sum_{d_1, d_2 = \pm1} \pi_3 \left(\sigma_1, \sigma_2, d_1; M^{(1)}\right) \pi_3 \left(\sigma_3, d_1, d_2; M^{(2)}\right) \nonumber \\
 && \ \ \ \ \ \ \ \ \ \ \ \ \times\pi_3 \left(\sigma_4, \sigma_5, d_2; M^{(2)}\right),
\end{eqnarray}
which is visually described in Fig. \ref{NspinTransformationStep}(a).
Repeating the DIT such that the DIT transformation order is as homogeneous as possible, we can show {\it a posteriori} that the general expression is given as
\begin{eqnarray}
\pi_p(\sset; M) = \Delta \sum_{{\bf d}} 
%\txtr{
\overbrace{\pi_3(\sigma_{1}, \sigma_{2}, d_{1}; M^{(n)}) \pi_3(\sigma_{3}, d_{1}, d_{2}; M^{(n)}) \cdots}^{2^{n+1}-(p-2)}
%}
 \nonumber \\
%\txtr{
\times \underbrace{\pi_3(\sigma_{p-1}, \sigma_{p}, d_{p-3}; M^{(n+1)}) \pi_3(\sigma_{p-2}, d_{p-3}, d_{p-4}; M^{(n+1)}) \cdots}_{2(p-2-2^n)},
%}
  \nonumber \\
\ 
\end{eqnarray}
where $n$ is an integer satisfying $2^n \leq p-2 < 2^{n+1}$. Note that there are two factors consisting of two visible spins, whereas the others contain only one. The number of factors with $M^{(n+1)}$ is zero if $p-2 = 2^{n}$.

To consider the hidden spins, we substitute 
%\txtr{
$M^{(n)}$
%}
 in the transformation introduced in Sec. \ref{3spinAsRBMSection} as, for instance,
\begin{eqnarray}
\pi_3(\sigma_1, \sigma_2, d_1; 
%\txtr{
M^{(n)}
%}
) = \Delta \sum_{\hset} \exp [(Wh + a)(\sigma_1 + \sigma_2 + d_1) + bh] \nonumber \\
\times \exp[W'[(\sigma_1 - \sigma_2)h_1 + (d_1 - \sigma_1)h_2 + (\sigma_2 - d_1)h_3]], \nonumber \\
\ 
\end{eqnarray}
which is described in the right-most panels of Fig. \ref{NspinTransformationStep}.

\section{Block Gibbs sampling from RBM equivalent to square Ising model}\label{GibbsSection}
In the following, we shortly introduce the block Gibbs sampling and discuss its application to the RBM equivalent to the ferromagnetic Ising model on the square lattice.
Let $A$ and $B$ be the sublattices of the square lattice. The spin configuration on a sublattice is denoted as $\sigma_{A(B)} = \left\{\{\sigma_i\}|i \in V_{A(B)} \right\}$ where $V_{A(B)}$ is the set of $A$($B$)-sublattice sites.
The posterior distribution for spin configuration on the $A$ sublattice is written as
\begin{eqnarray}
p(\sigma_A|\sigma_B) &:=& \pi(\sigma_A, \sigma_B)/\sum_{\sigma_A}\pi(\sigma_A, \sigma_B), \label{Hidden_Distribution} \\
\pi(\sigma_A, \sigma_B) &=& \exp\left(\beta \sum_{\langle i,j \rangle} \sigma_i \sigma_j\right),
\end{eqnarray}
where $\pi(\sigma_A, \sigma_B)$ denotes the Boltzmann factor for the total system and $\langle i,j \rangle$ denotes the edge connecting the sites $i$ and $j$.
Since spins in the $A$ sublattice do not couple to each other, Eq. (\ref{Hidden_Distribution}) can be factorized as
\begin{eqnarray}
p(\sigma_A|\sigma_B) &=& \prod_{i \in V_A} p(\sigma_i|\sigma_B),  \label{postProbFactorization}\\
p(\sigma_i|\sigma_B) &=& \frac{\exp\left[\beta \sigma_i \sum_{j \in \partial i}\sigma_j\right]}{2\cosh\left[\beta \sum_{j \in \partial i}\sigma_j\right]},
\end{eqnarray}
where $\partial i$ denotes the set of sites adjacent to $i$.
This allows us to sample each spin on $A$ sublattice independently without rejection. 
Update method that alternately switches the sublattice is called the block Gibbs sampling.

However, such an algorithm is not beneficial for the following reason. 
Consider a domain consisting of upward spins. 
In the bulk region of the domain, it is highly probable according to the Eq. (\ref{postProbFactorization}) that the newly sampled spins remain upwards as well. 
In other words, the limited range of the connection results in the mutual locking structure in the bulk region, allowing only the peripheral region to flip. 
Such a problematic situation is exacerbated as the domain size grows, and turns out that the slowing down is much worse than the single-spin flip.

\section{Partition function of three-spin interacting model on Kagom\'e lattice}\label{PartitionSection}
In this Appendix, we show that a model with only three-spin interactions on the Kagom\'e lattice is soluble. 
This model is a specific case of a broader class of models with crossing symmetry studied in Ref. \cite{Simon_Exactly_2013}, which do not exhibit a phase transition at finite temperature. The partition function is written as
\begin{eqnarray}
Z = \sum_{\sset}\exp \left[ \sum_{\up} M_{\up}\tau_{\up} + \sum_{\down} M_{\down}\tau_{\down}\right], \label{3spinPartition}
\end{eqnarray}
where $M_{\up}$ denotes the three-spin interaction and $\tau_{\up}:=\prod_{j \in \up} \sigma_j$ the product of the Ising spins in an upward triangle in the lattice. 
Also $M_{\down}$ and $\tau_{\down}$ are defined similarly for a downward triangle.
In order to compute Eq. (\ref{3spinPartition}), we introduce the identity for a binary variable $x = \pm1$ as follows,
\begin{eqnarray}
e^{Kx} = \cosh (K)\sum_{n=0,1}\left(x\  \tanh (K)\right)^n.
\end{eqnarray}
Applying this identity to each triangle yields
\begin{eqnarray}\label{3spinPartition2}
Z &=& \sum_{\sset} \sum_{n_{\up}, n_{\down}}
 \prod_{\up}\prod_{\down}\cosh (M_{\up})\cosh (M_{\down}) \nonumber \\
&&\ \ \ \ \ \ \times 
\left( \tau_{\up} \tanh (M_{\up})\right)^{n_{\up}} \left( \tau_{\down} \tanh( M_{\down})\right)^{n_{\down}}.
\end{eqnarray}

Next, let us consider taking the sum over $\sigma_j$ at some site $j$ in Eq. (\ref{3spinPartition}).
Denoting the triangles touching the site $j$ as $\up(j)$ and $\down(j)$, the contribution from the spin at $j$ can be given as 
$\sum_{\sigma_j} \sigma_j^{n_{\up(j)} + n_{\down(j)}}$, which is nonzero only when $n_{\up(j)} = n_{\down(j)}$.
This argument holds for arbitrary $j$, and therefore the requirement $n_{\up} = n_{\down} = 0, 1$ for all $\up$ and $\down$ imposed for nonzero contribution.
Accordingly, we obtain the concise expression of the partition function as
\begin{eqnarray}\label{3spinPartition3}
Z &=& C
\left(1 + \prod_{\up, \down}  \tanh (M_{\up}) \tanh (M_{\down})\right),
\end{eqnarray}
where $\displaystyle C = 
%\txtr{
2^{N_v}
%}
 \prod_{\up} \prod_{\down}\cosh (M_{\up}) \cosh (M_{\down}) $ 
%\txtb{
with $N_v$ being the number of visible spins, i.e., the number of sites.
%}
%\txtb{[$N_v$ is the number of visible spins $=$ the number of sites]}
The above expression clearly shows that the free energy in the thermodynamic limit is analytic, and hence the model does not show a phase transition at finite temperature.
%\txtr{
For $M_{\up} = M_{\down}$, Eq.~(\ref{3spinPartition3}) reproduces the partition function of the uniform model studied in Ref.~\cite{Muttalib17}. 
We note that the present method is not limited to two-dimensional models. 
In fact, a similar model with four-spin interactions on a three-dimensional pyrochlore lattice is also soluble using the same technique~\cite{Barry89}.
%}

\section{Cluster updates under magnetic field}\label{MagClusterSection}
In this Appendix, we introduce two flavors of cluster updates accompanied with magnetic fields. 
We assume a model defined on a graph $G = (V, E)$ as follows:
\begin{eqnarray}
H = -\sum_{\langle i,j \rangle \in E}J_{i,j} \sigma_i \sigma_j - \sum_{i \in V} h_i \sigma_i,
\end{eqnarray}
where $J_{i,j}$ is the two-spin interaction between two binary degrees of freedom at sites $i$ and $j$, or $\sigma_i$ and $\sigma_j$,
 and $h_i$ is the external field on site $i$. 
One way to take the external field into account is to modify the probability of flipping the cluster that is formed using the information on the interactions, and the other is to extend the space to express the field terms by interactions with the auxiliary space.

To introduce the first approach, let us remind that the detailed balance condition is given as,
\begin{eqnarray}
\frac{\pi(B)}{\pi(A)} =  \frac{P(A\rightarrow B) \alpha(A \rightarrow B)}{P(B\rightarrow A) \alpha(B \rightarrow A)},
\end{eqnarray}
where $\pi(A)$ is the Boltzmann weight corresponding to a state $A$,  $P(A \rightarrow B)$ is the trial proposal probability from the state $A$ to $B$, and $\alpha(A \rightarrow B)$ is the corresponding acceptance probability.
The Swendsen-Wang algorithm applied in the main text adopts the Metropolis-Hastings rule,
\begin{eqnarray}
\alpha(A \rightarrow B) = {\rm min}\left(1, \frac{\pi(B)}{\pi(A)}\frac{P(B \rightarrow A)}{P(A \rightarrow B)} \right) \label{AcceptanceRatio},
\end{eqnarray}
which satisfies $\alpha = 1$ under $h_i = 0$.
Since the external field modifies the Boltzmann weight as $\pi(A) \rightarrow e^{-\beta \sum_i h_i \sigma_i^{A}} \cdot \pi(A)$  at the  inverse temperature $\beta$,
the trial proposal must absorb such change to realize a rejection-free scheme under arbitrary external field.
Concretely, the $k$-th cluster $C_k$ formed by the ordinary bonding process is flipped with probability $p_k = e^{- \beta m_k}/(e^{- \beta m_k} + e^{+\beta m_k})$ where $m_k = -\sum_{i \in C_k} h_i\sigma_i$ is the Zeeman energy by the external field. The additional computational effort per single MC step is ignorable.

In the second approach, known as the ``ghost spin method,'' one introduces an auxiliary spin that interacts with any spin exposed to the external (or virtual) field \cite{Coniglio_Exact_1989, Dobias_Accelerating_2018}.
Defining $\widetilde{G} = (\widetilde{V}, \widetilde{E})$ with the ghost spin on the $0$-th site as
\begin{eqnarray}
V &\rightarrow& \widetilde{V} = \{0\} \cap V,\\
E &\rightarrow& \widetilde{E} = \{\langle 0, i\rangle | i \in V\},
\end{eqnarray}
we alternatively consider a Hamiltonian as follows,
\begin{eqnarray}
\widetilde{H} = - \sum_{\langle i,j \rangle \in \widetilde{E}}\widetilde{J}_{i,j} \sigma_i \sigma_j,
\end{eqnarray}
where 
\begin{eqnarray}
\widetilde{J}_{i,j} = 
\begin{cases}
J_{i,j}\text{\ \ \ if $\langle i,j\rangle \in E$}\\
h_i\text{\ \ \ if $j = 0$}\\
h_j\text{\ \ \ if $i = 0$}
\end{cases}.
\end{eqnarray}
Now that the new Hamiltonian consists solely of two-spin interactions, the ordinary cluster update can be applied.

\section{Observation of physical quantity in extended model}\label{ObservationSection}
The transformation considered in the main text preserves the partition function, and moreover the Boltzmann factor for visible spin configurations. 
Therefore, to compute the expectation value of a physical observable $O(\sset)$ in the extended space, one may simply consider the identical mapping $\widetilde{O}(\sigma, h) = O(\sigma)$ to obtain
\begin{eqnarray}
\langle O \rangle &=& \frac{\sum_{\sigma, h} \widetilde{O}(\sigma, h) \widetilde{\pi}(\sigma, h)}{\sum_{\sigma, h} \widetilde{\pi}(\sigma, h)} = \frac{\sum_{\sigma} O(\sigma) \left( \sum_h \widetilde{\pi}(\sigma,h)\right)}{Z}\nonumber \\
&=& \frac{\sum_{\sigma} O(\sigma) \pi(\sigma)}{Z}.
\end{eqnarray}
In other words, one may simply ignore all the hidden spins and compute the quantities using the operator in the original space.

\bibliographystyle{apsrev4-1}
\bibliography{rbm_mc_reference}

\begin{thebibliography}{71}
\bibitem [{com()}]{comment1}%
  \BibitemOpen
  \href@noop {} {\bibinfo  {journal} {Naively, it might seem that the mapping
  is inapplicable to frustrated systems due to the condition given in Eq.(12).
  However, we emphasize that any frustrated system can be mapped to RBM/DBM,
  for instance, by introducing virtual two-spin interactions}\ }\BibitemShut
  {NoStop}%
\end{thebibliography}

\end{document}